\pdfoutput=1 
\documentclass{article}
\usepackage{a4wide}
\usepackage[show]{ed}
\usepackage{xspace}
\usepackage{paralist}
\usepackage{mdframed}
\usepackage{multirow,rotating}
\usepackage{fancyvrb}
\usepackage{stex-logo}
\usepackage{local}
\usepackage[hyperref,backend=bibtex,style=alphabetic]{biblatex}
\renewbibmacro*{event+venue+date}{}
\renewbibmacro*{doi+eprint+url}{%
  \iftoggle{bbx:doi}
    {\printfield{doi}\iffieldundef{doi}{}{\clearfield{url}}}
    {}%
  \newunit\newblock
  \iftoggle{bbx:eprint}
    {\usebibmacro{eprint}}
    {}%
  \newunit\newblock
  \iftoggle{bbx:url}
    {\usebibmacro{url+urldate}}
    {}}
\def\pn{\textsf{Tetrapod}\xspace}

\newif\ifonly\onlyfalse

\usepackage{tikz}
\usepackage{tikzinput}
\usepackage{amssymb}

\usepackage{basics}
\usepackage{tetrapod}

\def\hateq{\ensuremath{\widehat=}\xspace}
\let\defemph\textbf
\def\svg{\textsf{SVG}\xspace}
\def\mathml{\textsf{MathML}\xspace}

\usepackage{hyperref}

\title{The Space of Mathematical Software Systems --- A Survey of Paradigmatic Systems}

\usepackage{authblk}
\author[1]{Katja Ber\v{c}i\v{c}}
\author[2]{Jacques Carette}
\author[2]{William M. Farmer}
\author[1]{Michael Kohlhase}
\author[1]{Dennis~M\"uller}
\author[1]{Florian Rabe}
\author[2]{Yasmine Sharoda}
\affil[1]{
Computer Science, 
FAU Erlangen-N\"urnberg, 
\url{http://kwarc.info}
\vspace*{1ex}}
\affil[2]{
Computing and Software, 
McMaster University,
\url{http://www.cas.mcmaster.ca/research/mathscheme/}
}

\begin{document}

\maketitle

\begin{abstract}
Mathematical software systems are becoming more and more important in pure and applied mathematics in order to deal with the complexity and scalability issues inherent in mathematics.
In the last decades we have seen a cambric explosion of increasingly powerful but also diverging systems.

To give researchers a guide to this space of systems, we devise a novel conceptualization of mathematical software that focuses on five aspects:
\emph{inference} covers formal logic and reasoning about mathematical statements via proofs and models, typically with strong emphasis on correctness;
\emph{computation} covers algorithms and software libraries for representing and manipulating mathematical objects, typically with strong emphasis on efficiency;
\emph{concretization} covers generating and maintaining collections of mathematical objects conforming to a certain pattern, typically with strong emphasis on complete enumeration;
\emph{narration} covers describing mathematical contexts and relations, typically with strong emphasis on human readability;
finally, \emph{organization} covers representing mathematical contexts and objects in machine-actionable formal languages, typically with strong emphasis on expressivity and system interoperability.

Despite broad agreement that an ideal system would seamlessly integrate all these aspects, research has diversified into families of highly specialized systems focusing on a single aspect and possibly partially integrating others, each with their own communities, challenges, and successes.
In this survey, we focus on the commonalities and differences of these systems from the perspective of a future multi-aspect system.

Our goal is to give new researchers, existing researchers from each of these communities, or outsiders like mathematicians a basic overview that enables them to match practical challenges to existing solutions, identify white spots in the software space, and to deepen the integration between systems and paradigms.

\smallskip\textbf{CAVEAT}: This paper is intended as a living survey that is updated in-place on \url{http://arXiv.org} from time to time. We publish this early as a pre-preprint to let the community discuss and maybe provide feedback to the authors at  \url{tetrapod@lists.informatik.uni-erlangen.de}. 
\end{abstract}


\newpage\tableofcontents\newpage

\section{Introduction}
In the last half decade we have seen mathematics tackle problems that lead to increasingly large developments: proofs, computations, data sets, and document collections. This trend has led to intense discussions about the nature of mathematics, ventilating questions like:
\begin{compactenum}[\em i\rm)]
\item Is a proof that can only be verified with the help of a computer still a mathematical proof?
\item Is a mathematical proofscape that exceeds what can be understood in detail by a single expert a legitimate justification of a mathematical result?
\item Can a collection of mathematics papers --- however big --- adequately represent a large body of mathematical knowledge?
\end{compactenum}
In~\cite{CarFarKohRab:bmobb19} we have discussed these questions under the heading of \defemph{Big Math} and propose a unified, high-level model.
We claim that computer support will be necessary for scaling mathematics, and that suitable and acceptable methods should be developed in a tight collaboration between mathematicians and computer scientists --- indeed such method development is already under way, but needs to become more comprehensive and integrative.

\begin{figure}[ht]\centering
\includegraphics[width=5cm]{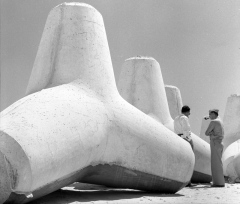}\qquad
\providecommand\myscale{4.5}
\begin{tikzpicture}[scale=\myscale]
  \node (center) at (0,.15) {Organization};
  \node (left) at (.2,-.3) {Computation};
  \node (right) at (.4,0) {Concretization};
  \node (back) at (-.5,0) {Inference};
  \node (up) at (0,.5) {Narration};

  \draw[very thick] (center) -- (left);
  \draw[very thick] (center) -- (right);
  \draw[very thick] (center) -- (back);
  \draw[very thick] (center) -- (up);
  \draw[dotted] (left) -- (right) -- (back) -- (left);
  \draw[dotted] (up) -- (left);
  \draw[dotted] (up) -- (right);
  \draw[dotted] (up) -- (back);
\end{tikzpicture}
\caption{Five Aspects of Big Math Systems, a Tetrapod Structure}\label{fig:tetrapod}
\end{figure}
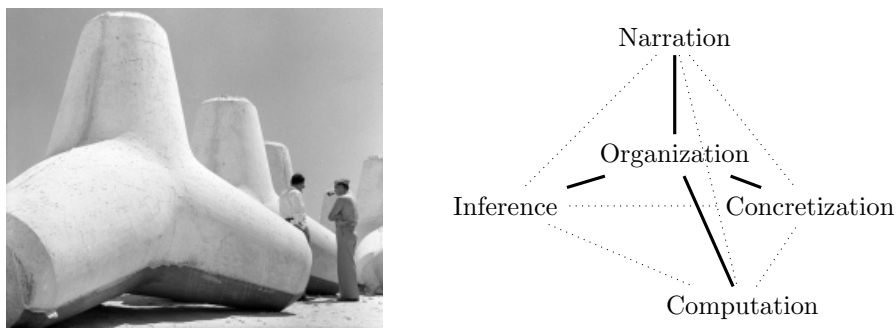

We propose that all Big Math developments comprise four main aspects that need to be dealt with at scale:
\begin{compactenum}[\em i\rm)]
\item \emph{Inference}: deriving statements by \emph{deduction} (i.e., proving), \emph{abduction} (i.e., conjecture formation from best explanations), and \emph{induction} (i.e., conjecture formation from examples).
\item \emph{Computation}: algorithmic manipulation and simplification of mathematical expressions and other representations of mathematical objects.
\item \emph{Concretization}: generating, collecting, maintaining, and accessing collections of examples that suggest patterns and relations and allow testing of conjectures.
\item \emph{Narration}: bringing the results into a form that can be digested by humans, usually in mathematical documents like articles, books, or preprints, that expose the ideas in natural language but also in diagrams, tables, and simulations.
\end{compactenum}
These aspects --- their existence and importance to mathematics --- should be rather uncontroversial.  Figure~\ref{fig:tetrapod} may help convey the part which is less discussed, and not less crucial: that they are tightly related.  For a convenient representation in three dimensions, we choose to locate the organization aspect at the barycentre of the other four since they are all consumers and producers of mathematical knowledge.

Computer support exists for all of these four aspects of Big Math, e.g.,
\begin{compactenum}[\em i\rm)]
\item theorem provers like Isabelle, Coq, or Mizar;
\item computer algebra systems like GAP, SageMath, Maple, or Mathematica; and
\item mathematical data bases like the L-functions and Modular Forms Data Base (LMFDB)~\cite{Cremona:LMFDB16,lmfdb:on} and the Online Encyclopedia of Integer Sequences (OEIS)~\cite{Sloane:OEIS};
\item online journals, mathematical information systems like zbMATH or MathSciNet, preprint servers like arXiv.org, or research-level help systems like MathOverflow.
\end{compactenum}
While humans can easily integrate these four aspects and do that for all mathematical developments (large or otherwise), much research is still necessary into how such an integration can be achieved in software systems.
\ednote{MK: Possible outline:
\begin{enumerate}
\item The mathematics process involves several integrated activities.
\item Contemporary mathematical software systems usually focus on just one of these activities.  Moreover, these systems are not designed to work with each other and do not employ a common knowledge base.
\item The limitations of contemporary systems have been exposed by the tremendous growth in the production of mathematical knowledge and ``big math'' projects like the Classification of Finite Simple Groups.
\item We need a mathematical software system that is holistic in the sense that it is designed to support the entire mathematics process and to be integrated with existing systems.
\item The first step towards the goal of a holistic system is to survey the mathematical software systems that are available today, which is subject of this paper.
\end{enumerate}}

\paragraph{Overview} 
We want to throw the spotlight on the integration problem to help start off research and development of systems that integrate all four aspects.
To facilitate this, we give a high-level survey of mathematical software systems from the \pn perspective.
Because almost every one of these systems has one primary aspect and because systems with the same primary aspect are often very similar, we group systems by their primary aspects.

In each group we try to further subdivide the systems.
We want to stress that this classification is mostly meant for convenience, e.g., by simplifying the description of several similar systems.
We do not mean to imply a strict separation between these groups of systems, and often the borders are fluid.
In particular, there are some systems that already allow aspect switching in a way that precludes ascribing a primary aspect.
For convenience, we still assign these systems into one of the groups and discuss which
other aspects they support.\ednote{MK: I do not think we should do it that way, but
  mention them in both}

While most systems use one of the four aspects as the primary one, there are also some systems that focus primarily or even exclusively on the ontology.
Therefore, use a fifth group\ednote{MK: this is currently the first one discussed, we either need to announce that or move it last.} for those systems.

In survey in the next sections we use the following general terms for all aspects:
\emph{Syntax} is a set of rules for forming objects, and \emph{data} is any piece of well-formed syntax.
\emph{Semantics} is a set of rules for interpreting objects, and \emph{knowledge} is a pair of a datum and its semantics.



\section{Primary Aspect: Inference}\label{pa:inf}
Various methods have been developed to represent and perform inferences.
We structure our presentation by how each method relates to computation, the aspect most whose integration with inference has drawn the most attention.
In general, the ubiquity of underspecified function symbols and quantified variables means that logical expressions usually do not normalize to unique values.
At best, computations like $y:=f(x)$ can be represented as open-ended conjectures where different options for $y$ are produced, each together with a proof of the respective equality.
Therefore, inference systems usually sacrifice computation or at least its efficiency.

\emph{Proof assistants} sit at the extreme end of this spectrum.
They employ strong logics and high-level declarations to provide a convenient way to formalize domain knowledge and reason about it.
The reasoning is usually interactive in order to represent inferences that are too difficult to be fully automated.
Most proof assistants integrate at least some of the other methods to overcome this weakness.

Further along the spectrum, \emph{automated theorem provers} use simpler logics than interactive proof assistants.
They are fully automatic and much faster, but can handle much fewer theorems, and typically do not check their proofs.
\emph{Satisfiability checkers} continue this progression by aiming at decidable automation support, whereas theorem proving is usually an semi-decidable search problem.
That limits them to propositional logic or specific theories of more expressive logics (usually of first-order logic) that are complete, i.e., where every formula can be proved or disproved.
In the special cases, where satisfiability checkers are applicable, they come close to verified computation systems.

Orthogonal to the above triplet, there are several methods for realizing Turing-complete computation naturally inside a logic.
Here imperative and object-oriented computation are usually avoided in favor of other programming paradigms that are easier to reason about.
\emph{Rewriting} aims at optimizing the $f(x)\rewrites y$ progression, allowing users to mark specific transformations as rewrite steps.
\emph{Terminating recursion} is the method of adding recursive functions to a logic in order to make it a pure functional programming language.
Finally, \emph{logic programming} restricts attention to theorems of a special form, for which proof search is simple and predictable so that users can represent computations by supplying axioms that guide the proof search.

In the sequel, we describe each method in some more detail.

\paragraph{Proof Assistants (C)}
The most successful proof assistants represent tens or hundreds of person-year investments into
\begin{compactitem}
 \item Define the foundational logic.
 Usually logics much stronger than textbook first- or higher-order logic are needed in practice.
 A big question has been the trade-off between flexible, untyped languages that rely on undecidable reasoning and rich type system that are more restrictive but have better computational properties.
 Typical choices are first-order set theory (e.g., Mizar \cite{mizar}), higher-order logic (e.g., HOL \cite{hol4}, HOL Light \cite{hollight}, Isabelle/HOL \cite{isabelle}, PVS \cite{pvs}), and constructive type theory (e.g., Coq \cite{coq}, Matita \cite{matita}, Lean \cite{lean}, Agda \Cite{agda}).
 Some systems use undecidable type systems (e.g., Mizar, PVS) as compromises. 
 \item Implement the logic.
 Usually the implementation starts with a kernel that checks proofs and then grows outwards in layers until a human-friendly surface syntax is exposed.
 Much work has been put into automatically filling in as many gaps left by the user as possible.
 Tactic languages (e.g., HOL, HOL Light, Coq, Isabelle) and high-level proof languages (e.g., Isabelle, Coq) and the integration of automated provers (e.g., Isabelle, Coq) and decision procedures (e.g., PVS, Isabelle) have been crucial here.
 \item Build a library of data structures.
 Usually proof assistants are only valuable if their standard library provides many data basic structures of mathematics (e.g., groups, real function) and computer science (e.g., records, inductive types).
 This wa most prominently envisioned in the QED manifesto \cite{qed}.
 But representing these and proving their characteristic properties has proved very expensive and remains a tough benchmark for the design of logic and system.
\end{compactitem}

The tactic languages are often Turing-complete themselves (e.g., HOL Light, Isabelle, Coq).
This is realized by writing tactics in the underlying programming language.
Recently systems have tried to represent tactics in the systems itself either declaratively (e.g., Coq, Matita) or programmatically through reflection of the kernel data structures (e.g., Lean).

Most major proof assistants allow interspersed narrative structure, at the very least through comments.
Some systems mimic sectioning where the scope of variables is determined not by the logic but by the narrative structure (e.g., Coq).
Some systems (e.g., Isabelle, Agda) are narratively strong enough to make it feasible to write narrative documents (most importantly the documentation of the system itself) in the system.

\ednote{JC: recent paper "A Survey on Theorem Provers in Formal Methods" -
https://arxiv.org/abs/1912.03028 has just appeared. We should at the very least cite
all the software that appears in it.}

%

\paragraph{Automated Theorem Provers (A)}
Automated provers have been mostly developed for relatively simpler logics, where full automation is feasible.
Therefore, they are often used as backend system integrated into, e.g., interactive proof assistants.

Most systems work with variants of first-order logic and compete regularly in the CASC competition.
Examples are Vampire \cite{vampire}, E \cite{e}, and Spass \cite{spass}.
An ongoing trend is the gradual extension of first-order logic with additional features such as definitions, primitive numbers, types, and polymorphism.

Automated provers for higher-order logics are much harder to develop but are gaining strength.
Example systems are Leo \cite{leo2} and Satallax \cite{satallax}.

Recently efforts have been made for automated provers to return checkable proofs in order to use them in proof assistants (see \cite{qedhammer} for an overview).
A big problem here is the selection of useful axioms to reduce the search space for the automated prover.
Recently machine learning has been employed successfully for this purpose (e.g., \cite{holyhammer}).

\paragraph{Satisfiability Checkers (S)}
While satisfiability is decidable for propositional logic and a few first-order theories (most importantly variants of linear arithmetic), the overall scope is limited.
Decision problems are typically stated in the form of satisfiability problems and called \emph{SAT solvers} (for propositional logic) or \emph{SMT solvers} (satisfiability modulo theory, for specific theories of, typically, first-order logic).
Specific decision procedure-based systems for higher logics include Z3 \cite{z3} and CVC \cite{cvc}.
\ednote{FR: cite some SAT and SMT solvers here}

Satisfiability checkers for propositional logic have become so powerful that it is handle feasible to encode high-level problems as propositional problems by using large amounts of propositional variables.

To increase scope, efforts are made at combining decision procedures or applying them to theories with decidable fragments.
While these efforts lose decidability, they may still be highly valuable in practice, e.g., by directly computing a value instead of extracting it from a proof, or to reduce the search space by eliminating unsatisfiable branches.

\paragraph{Terminating Recursions (T)}
Many proof assistants include sound and incomplete termination checkers (e.g., PVS, Coq, Agda) that only accept provably terminating functions.
These are Turing-complete in the sense that the syntax can represent all computable functions but the termination checker will not accept all of them.
Some systems (e.g., Isabelle, Coq) can export those programs in external programming languages after verifying their correctness in the logics.

In some cases it can be hard to draw the line between inference and computation systems.
This is the case for systems that use functional programming combined with the representation of proofs-as-programs (e.g., Coq, Agda).

\paragraph{Rewriting (R)}
Rewriting implements a directed equality relation between expressions.
This can be used both to obtain Turing-complete computation and for reasoning (by rewriting formulas to a boolean value).

Compared to other forms of computation, rewriting is very inefficient, e.g., rewriting polynomials into normal form takes exponentially longer than an algorithm based on plain ``arithmetic''.
But the embedding of rewrite systems in logic permits proving the soundness of each rewrite rule.

Originally most rewrite systems were based on first-order logic due to its decidable unification (e.g., Maude \cite{maude}).\ednote{add more citations}
But recently more systems are employing rewriting in higher-order setting (e.g., Dedukti \cite{dedukti}). \ednote{check Cynthia Kop's talk on higher-order rewriting for more citations}

A lot of effort has gone into establishing the confluence and termination of sets of rewrite rules.
This is critical to establish that rewriting implements a deterministic computation.
Therefore, many systems try to establish confluence and termination of sets of rewrite rules automatically. \ednote{cite examples}

Many proof assistants integrate rewrite engines that use some of the proved theorems as rewrite rules that are applied automatically by the system (e.g., PVS, Isabelle).

\paragraph{Logic Programming (L)}
If the proof search behavior of a prover is known to the user and predictable in practice, this can be instrumented for computation.
This works particularly well with axioms in Horn-form where searching a proof of the conclusion triggers searching proofs of the assumptions.
This is the basic idea of logic programming, where a formalization consists of a set of Horn axioms, which can be seen as both a specification and (via predicatble proof search) as a program.
If special predicate symbols are added, whose ``proofs'' consists of extra-logical operations like I/O side effects, this yields a general purpose programming language.

Logic programming was mostly investigated in untyped first-order logic via various Prolog dialects \cite{ClMe81}.
But higher-order variants exists as well, e.g., such as $\lambda$Prolog \cite{lambdaprolog}.
\ednote{need more references here}


\ednote{Responsible: Florian}

\section{Primary Aspect: Computation}\label{pa:comp}
\ednote{Responsible: Jacques}
\ednote{MK: We need letters for the big table here. I will use ``T'' for Turing
Complete; need that for \TeX}

Computation is a rather broad topic: for example, term rewriting systems
and finite state machines are often regarded as performing computations.
At the Turing-complete end of the spectrum, we could list all programming
languages as performing computations --- because, well, they do!

As our primary focus is still centered on mathematics, this helpfully
narrows things down. Even though one can indeed use just about any language
to do mathematics, it makes sense to instead focus on those languages
that have been designed with mathematics in mind.

The systems can be usefully divided according to the kinds of data they
were primarily designed to handle:
\begin{itemize}
\item Typed
\item Symbolic
\item Algebraic
\item Numeric
\begin{itemize}
\item analytic
\item statistical
\end{itemize}
\end{itemize}

By \emph{typed} data, we mean data that can be expressed in ``type theory'',
as originating from Russell and Whitehead~\cite{WhiRus:pm10}, through
Church~\cite{Church:afotst40}, until today; the historical development until
$10$ years ago is well documented in~\cite{kamar}.  Although we have seen
a wide varieties of type theories used for systems that perform inference,
for systems that take computations seriously, one family emerges:
dependent type theories.  By \emph{symbolic} data, we mean data that can
be best expressed as \emph{abstract syntax trees}, usually containing
``free variables''. From a type-theoretical point of view, these are
significantly harder to deal with, as these would then form \emph{open terms},
which are notoriously difficult to manipulate correctly. This is why all
know symbolic computation systems are untyped.  By \emph{algebraic}, we
basically mean data that belongs to the mathematical sub-domain of
Algebra. What distinguishes these is that, although it is frequently
convenient to use ``free variables'' for the visual display of these objects,
they are not fundamentally required for an adequate representation.  Under
Algebra, we also include systems that do exact computations on natural
numbers, integers, Gaussian integers, etc, as these are also algebraic.
By \emph{numeric}, we mean systems whose data include ``real numbers'' in
one way or another; it is useful to subdivide this class further, into the
systems that specialize in more analytic problems (quadrature, differential
equations) from those that deal with statistics. Both kinds excell at
computations based on linear algebra.

Some of the systems that we survey below are quite broad, and so implement
many features in common: one can indeed do statistics in Maple, and symbolic
computation in Matlab. The classification is not meant to ``squeeze'' any
system into a narrow box, but rather to express the \emph{fundamental
organizational system} around which the system grew outward to encompass
much more.

\paragraph{Typed}

Agda \cite{agda}, Idris \cite{idris}

ATS.

\paragraph{Symbolic}
Mathematica

Maple

Axiom

\paragraph{Exact Computation Systems}
Gap: groups

Singular: ideals

Sage: integrated with Python and (via Python) other languages

\paragraph{Scientific Computation}
Matlab, scilab, Octave.

ChebFun, NumPY,

\paragraph{Statistics Packages}

R. SPSS, SAS, Minitab.

\paragraph{More categories}

Machine learning, probabilistic programming.  Term rewriting?

\paragraph{commentary}
\ednote{while I understand why this was put here, I think this should just
be deleted. In many ways, this is anti-tetrapodian thinking, as it 
priviledges inference over everything else.}
By \emph{reflection}, we mean the ability of a logic to have access to (some
of) the reasoning facilities usually associated with its meta-logic.
Computational reflection is similar, in the context of programming languages;
this usually proceeds via a representation of some (otherwise opaque) concepts,
which can be manipulated, before being \emph{reified}.  If the  representation
is adequate and the manipulations are meaning preserving, then computation can
implement reasoning.

If programming languages use a sufficiently strong \emph{type system}, they may
be able to embed deduction into computation.  Types are, via Curry-Howard,
(simple) propositions that are considered true if they are inhabited.  Then
type checking is a form of verification (i.e., checking that a particular
proof/term inhabits a particular proposition/type).  \emph{Abstract
interpretation} moves beyond ``simple'' types into being able to attach
significantly more powerful properties to programs.



\section{Primary Aspect: Concretization}\label{pa:tab}
\ednote{responsible: use summary of Katja's survey}
\paragraph{Overview}
The naming of this aspect of knowledge section has proved surprisingly difficult.
We mean to include any practice of representing mathematical objects in terms of concrete data structures.
Mathematics has a long tradition of such efforts, going back to, e.g., clay tablets of Pythagorean triples~\cite{Abdulaziz:plimpton10}, lists of decimal digits of $\pi$, or logarithm tables.
More modern incarnations include computer-supported practices like large prime numbers, the database of integer sequences, and the enumeration of isomorphism classes of simple finite groups.
But this practice does not a standard name.

By \emph{concrete data structures}, we mean any data formed using primitive objects such as integers and strings via constructors like lists, records, and tables.
These objects have in common that they have an objective physical reality that is beyond doubt: for example, any particular finite list of integers exists absolutely, whereas the existence of, e.g., proofs or programs may be relative to philosophical assumptions (e.g., impredicativity, classical reasoning, axiom of choice) or mathematical conditions (e.g., soundness of an argument, termination of an algorithm).
Thus, our concrete objects are tangible and material in a way that is opposite to the Platonic objects that deduction and narration and to some extent computation are concerned with.
Thus, we can also think of them as the shadows of Platonic philosophy or as Aristotelian objects in the sense of being empirical, observable, and practical.

Most of the time, concrete objects are aggregated in the form of tables such as logarithm tables or the many large relational databases described below.
We considered \emph{tabulation} as an alternative name for this aspect but opted against it to avoid excluding other representation languages for concrete data such as arrays and JSON.

\paragraph{Encodings}
There are two ways to define the semantics of concrete objects.
Firstly, we can characterize the concrete objects as the subset of those mathematical objects that are self-denoting: e.g., any natural number is interpreted as itself.
This is in contrast to the other three aspects, where interpretations are needed to map objects to their denotation.
Secondly, many non-concrete mathematical objects can be represented in terms of concrete ones, a process that we call \emph{encoding}.
Encodings are commonly used in databases of algebraic structures such as elliptic curves or isomorphism classes of graphs.
Seen mathematically, any encoding is based on a representation theorem, which states the encoding $e(o)$ of an object $o$ fully characterizes $o$ (up to isomorphism ideally).
However, such representation theorems do not always exist because sets and functions, which are the foundation of most mathematics, are inherently hard to represent concretely.

Conceptually, any effective representation of mathematical objects requires some encoding as concrete objects because only those can be acted on by computers.
Therefore, in general, the semantics of concrete objects consists in applying the dual \emph{decoding} operation.
We speak of \emph{codecs} for a pair of encoding or decoding that represent a class of mathematical objects as concrete ones.

\paragraph{Data and Knowledge}

Until the advent of computer-supported mathematics, the creation and sharing of data only received attention in passing.
The logarithm tables are a good example of this, as are the examples 
collected in the community effort initiated by Gordon Royle on MathOverflow~\cite{Royle:list:on}.

Billey and Tenner introduced the concept of a fingerprint database of theorems in 2013~\cite{BilTen:fingerprint13}.
The primary example of this is the OEIS.
They stress the importance of the following aspects: 
searchability, collaborativeness, citability of the contents, and indexing by small, language independent and canonical data.

Mathematical datasets and databases today are highly varied 
and range from small to large in several aspects.
The datasets can easily reach the Gigabyte range: 
LMFDB (~1TB data in number theory),
or a lattice dataset by Kohonen (uncompressed about 1.5TB of lattices),
to name some of the largest.
The can range from short lists of objects that are extremely hard to compute, to
gigantic lists of millions of objects, such as the GAP Small Groups Library (about 450 million finite groups)
and the previously mentioned dataset of $17 \cdot 10^9$ lattices.
Similarly, the authorship varies from single-author datasets, to community efforts such as the OEIS,
with thousands of contributors.
The structure of mathematical datasets can be as simple as having a list of objects,
a list of records that can contain information like mathematical invariants in addition to the object,
to complex databases of related tables such as the LMFDB.

Commonly occurring themes in mathematical data are ad-hoc implementations of codecs,
lack of community guidelines for data, and similar.\ednote{Mention FAIRMat?}

We will use the following facets for classifying systems in Table~\ref{bigtable}: We start out with three classes of representations of concrete objects: 

\newAspectFacet[C]{Record Data}R{} where datasets are sets of records conforming to the same schema.  Record data and querying is very well-standardized by the relational (SQL) model.  However, if encodings are used, SQL can never answer queries about the semantics of the original object.

\newAspectFacet[C]{Array Data}A{} consists of very large, multidimensional arrays that require optimized management.  Array data tends to come up in settings with large but simply-structured datasets such as simulation time series, while record data is often needed to represent complex objects, especially those from pure mathematics.  Array data bases, which offer efficient access to contiguous --- possibly lower-dimensional --- sub-arrays of datasets (voxels), are less standardized, but OPenNDAP~\cite{OPenNDAP:on} is becoming increasingly recognized even outside the GeoData community, where it originated.

\newAspectFacet[C]{Linked Data}L{} introduces identifiers for objects and then treats them as blackboxes, only representing the identi- fier and not the original object. The internal structure and the semantics of the object remain unspecified except for maintaining a set of named relations and attributions for these identifiers. The named relations allow forming large networks of objects, and the attributions of concrete values provide limited information about each one. Linked data can be subdivided into knowledge graphs and metadata, e.g., as used in publication indexing services.

\newAspectFacet[C]{Isomorphism Classes}I{}
In many cases, all mathematical properties of an class of objectsare invariant under the natural isomorphisms of the class.
In this case, we are less interested in the objects themselves, but only in their isomorphism classes.
Groups, graphs, and elliptic curves are prime examples of this.
Some math data systems have sustematic representing isomoprhismclasses, which can be considerably more difficult to represent than their representatives.
To achieve this e.g. for graphs we need to generate a canonical labeling for the encoding, i.e. where two graphs are isomorphic, whenever the canonical labeling is identical.
And then we need to port all the persistence layer, the algorithms, and the UI to the enhanced graph encoding. 

\newAspectFacet[C]{Redundant Information/Computed Properties}P{} Often, a small set of properties of an object suffice to characterize if fully (up to isomorphism); we will call such properties constitutive.  Some concretization systems\ednote{MK: Do we really want to say ``concretization system'', what could we say better?} support the storage of ``redundant information'', i.e. information about of the objects that can in principle be computed from the constitutive properties, but may be too costly, or might be needed as keys for object selection queries. 

\newAspectFacet[C]{Complete Enumeration}E{}
Contrary to most other forms of represented mathematical knowledge, concrete mathematical data provides the chance to completely enumerate a set (of isomoprhism classes)  of objects. E.g. all groups up to a given order. In all other collections of represented mathematical knowledge -- think theorem prover libraries, preprint collections, or computer algebra systems -- objects are curated because they are ``interesting'' because they have special properties. 

\newAspectFacet[C]{Named Objects/Symbolic Terms}S{}\ednote{MK@FR: I am not sure what to say here. }

\newAspectFacet[C]{Integration with Computation System}C{}
Most mathematical software systems that store and manage represented mathematical knowledge are based on a  conventional persistence layer, i.e. files with object encodings, a relational, array, or graph data base together with a user interface.
The knowledge itself is usually created by users utilizing the system itself -- e.g. proofs created in the system, papers written in {\LaTeX}, or programs developed developed in the integrated IDE -- and are curated by a community or a commercial entity. 
Alternatively -- especially in the cases of complete enumeration (see above) -- the content can be created programmatically by a computational system.
In this case, there is usually a tight integration with a dedicated computational system. 
\begin{oldpart}{not clear if we need this. }
\paragraph{Standalone Databases}


OEIS \cite{OEIS:on}: some ontology and narration

LMFDB \cite{lmfdb:on}: content produced by computation; filled by computation; computation integrated into frontend via Sage; narration and ontology for background knowledge

FindStat \cite{findstat}: 

ATLAS of Finite Group Representations (\url{http://for.mat.bham.ac.uk/atlas/}):

Database of Ring Theory (\url{http://ringtheory.herokuapp.com/}):

Math Counterexamples (\url{http://www.mathcounterexamples.net/}):

Manifold Atlas: (\url{http://www.map.mpim-bonn.mpg.de})

Distributome (\url{http://www.distributome.org/})

SuiteSparse (\url{https://sparse.tamu.edu/})

House of Graphs (\url{http://hog.grinvin.org})

Digital Library of Mathematical Functions (\url{https://dlmf.nist.gov/})

\paragraph{Data Sets within Computation Systems}

Mathematics (embedded and external data sources)

Sage

GAP data libraries (\url{http://www.gap-system.org/Datalib/datalib.html}):
(e.g., table of transitive groups)
\end{oldpart}


\section{Primary Aspect: Narration}\label{pa:nar}
\ednote{responsible: Michael}
The ``narration'' aspect of the tetrapod is concerned with mathematical knowledge in a form that can be digested by humans.
Narratively represented mathematical knowledge usually exists in mathematical documents\footnote{While we take a rather inclusive view on ``documents'', we limit ourselves to written documents, excluding audio recordings and videos. Our rationale is that this does not constitute a loss of generality since the latter could be transcribed into written documents without significant loss in meaning.} like articles, books, or preprints, that expose the ideas in natural language but also in diagrams, tables, and simulations.
While rigour and correctness are important concerns in narration, the main emphasis is on communicating ideas, insights, intuitions, and inherent connections efficiently to colleagues well-versed in the particular topic or students who want to become that.
As a consequence, more than half of the text of a typical mathematical document consists of introductions, motivations, recaps, remarks, outlooks, conclusions, and references.
Even though the ``packaging'' of mathematical knowledge into documents leads to some duplication in the mathematical literature, it seems to be an efficient way of dealing with communication and knowledge preservation and can thus be seen as a necessary overhead in scholarly communication.

The primary aim of this survey is to explore holistic mathematical software systems --  here software support for mathematical documents.
This in turn depends on the depth of explicit structural and semantic markup in documents.
Currently, there are four (plus one) levels of representation of mathematical documents:
\begin{compactenum}
\item[RL0.] \defemph{written up}: for communication (e.g. chalk on blackboards) or archival purposes -- e.g. on papyrus scrolls \hateq $\sim$90\%\footnote{The percentages in this classification are rough estimates that should be taken as qualitative indications of the relative size rather than actual quantities.} of the mathematical documents.
\item[RL1.] \defemph{digital} usually digitized from print -- e.g. as TIFFs \hateq $\sim$50\%
\item[RL2.] \defemph{presentational}: encoded text interspersed with presentation markup -- e.g. PDF, Word, {\TeX} or presentation \mathml \hateq $\sim$20\%
\item[RL3.] \defemph{semantic}: encoded text with functional markup for the meaning, e.g.  {\LaTeX}, {\sTeX}, Mathematica Notebooks \hateq $\sim$1\%
\item[RL4.] \defemph{formal}: The meaning of the document is fully specified and thus machine-actionable at all levels. $\leq$0.1\%
\end{compactenum}
We remark that the delineation of levels is somewhat fuzzy and that the ``levels'' themselves are far from uniform.
Nevertheless they constitute a useful categorization for our discussion.   
Especially in the higher levels, the presentational and semantic markup is  usually restricted to particular aspects of the document functionality.
We have ordered the examples of representation formats by increasing opportunities for structural and functional markup.

Note transforming from higher levels to lower levels is usually simple via largely context-free \defemph{styling} rules, whereas the opposite direction -- \defemph{semantics extraction} --  involves non-trivial context-dependent heuristic choices.
For instance in the transformation between the levels RL1. and RL2. we have the difference between printing (down transformation) or OCR (optical character recognition; up transformation). 

We finally note that computation and thus machine support needs explicitly represented structures and thus higher representation levels lead to more opportunities for software support.
Therefore we will structure our survey bottom-up in the hierarchy above, starting with level RL2. since levels RL0. and RL1. have no discernable math-specific aspects. 

\newAspectFacet[N]{At the Presentational Level}P{Word Processors \& Document Preparation Systems}

At this level, we have any kind of software system and for document preparation that can deal with \defemph{mathematical vernacular}, the peculiar mixture of natural language, mathematical formulae, and diagrams digitally.
In contrast to the image-based formats at level RL1., text is encoded as sequences of characters, whereas formulae and diagrams are in some kind of presentational markup.  This is sufficient to e.g. make documents at the presentation level searchable in conventional bag-of-words-based search engines like Google or Bing.
The systems and representation formats mainly differ in their 
\begin{compactenum}
\item \defemph{authoring model}: WYSIWYG or formatted/programmable text,
\item \defemph{target media}: paginated or flexible page size, 
\item treatment of mathematical formulae.
\end{compactenum}
Word processors like \textsf{MS Word} or \textsf{LibreOffice Writer} implement a WYSIWYG -- ``what you see is what you get'' -- authoring model and target paginated media.
Typesetting systems like {\TeX/\LaTeX}, the document preparation system preferred in mathematics, let authors encode documents as Unicode strings with executable -- often user-definable --  control sequences, which a formatter expands into a primitive page description format.
For publication, the standard target page description format in both cases is usually PDF (Adobe's standardized Portable Document Format).
This fixes page layouts down to the character position.
Diagrams and formula components that do not come from one of the available fonts are represented as vector graphics.
In particular, mathematical formulae lose all structural information during the PDF transformation, so that higher-level services like mathematical search or screen readers have nothing to go on, and would OCR-like facilities to function. 
In essence, mathematical  formulae and diagrams are still at level RL1 (image-like)  in PDF, even though the ``source'' (\textsf{Office Open XML}~\cite{OOXML:spec:on} for \textsf{MS Word} and \textsf{Open Document Format}~\cite{DurBrau:odfoa11} for \textsf{LibreOffice Writer}) may still have had the necessary structures.

Most document preparation systems also allow the export of \textsf{HTML5}~\cite{W3C:html5.2}, a web markup format for interactive multimedia documents, that can encode mathematical formulae via \mathml (the Mathematical Markup Language; see~\cite{CarlisleEd:MathML3:on}) and diagrams via \svg (Scalable Vector Graphics; see \cite{W3C:svg11}).
For \textsf{MS Word}, \mathml export is a native feature, for \textsf{LibreOffice Writer} via a plugin \textsf{Writer2xhtml}~\cite{writer2xhtml:on}, {\TeX/\LaTeX} can be exported to \textsf{HTML5} (and the eBook format \textsf{ePUB3}~\cite{EPUB3} based on it) via the {\LaTeX}ML engine or {\TeX4HT}. 
Note that \mathml has two sub-languages: \defemph{presentation \mathml} specifies the visual layout of formulae (i.e. level RL2), and \defemph{content \mathml} for the meaning (the associated operator trees; i.e. RL3).
The exports above all restrict themselves to presentation \mathml (though {\LaTeX}ML~\cite{Miller:latexml:online} does a best-effort attempt at inferring content \mathml).
But even presentation \mathml has enough structure to support formula screen readers like MathPlayer~\cite{mathplayer} or mathematical search engines; see~\cite{GuiSac:srmk15,AizKohOunSch:nmto16} for pointers to the state of the art and~\cite{zbf:on} for an online example.

\newAspectFacet[N]{Formal, Narrative Documents}F{}

To understand the semantic representation level (RL3) of narrative mathematical documents, which mixes language and with formal aspects, it is good to think about fully formal narrative documents.
By definition, these could consist of a sequence of logical propositions\footnote{In fact, the resolution of the Grundlagen Crisis of Mathematics in the last  century is that any mathematical document can -- in principle -- be formalized in first-order logic with axiomatic set theory.}, possibly extended by a formalization of sectioning, discourse, and rhetorical structures.
To the best of the knowledge of the authors, there are no fully formal, narrative representations of mathematical knowledge.
All fully formal representations (e.g. proof assistant libraries cf. Section~\ref{pa:inf}) are organized with respect to the inherent structures of the knowledge space, with little concern to a given narrative for human readers. 
Where  attempts of a narrative are made -- e.g. in comments -- these are informal natural language, not logic.    
A notable example is van Benthem Jutting' s formalization \cite{vBJ:clgas77} of Landau's ``\emph{Grundlagen der Analysis}~\cite{Landau30}, where the formalization closely follows the structure of the original.
Arguably the narrative structure of the narrative structure of the Grundlagen is very limited, and the formalization excluded the explanatory introduction.
The NaProChe~\cite{CramerFKKSV09,MKM11:PDSMNS} project develops a controlled natural language for mathematics, i.e. a formal language that is -- syntactically -- a subset of mathematical vernacular, with the aim of verifying mathematics in the \textsf{Isabelle} proof assistant.
Again, introductions, motivations, recaps, remarks, outlooks, conclusions, and references are not part of the language. 

A similar but dual example -- verbalization instead of formalization -- is the case of Mizar articles, from which human-oriented presentations for the Journal of Formalized Mathematics~\cite{JFM:on} are generated.
Again, abstracts and introductions have to supplied by (human) authors.
The remaining content is generated from the Mizar theorems and proofs. The latter are sequences of statements and justifications which can be verified by the Mizar prover.
In fact, formal proofs given as proof step sequences may be the only fully formal narrative documents currently available.
In contrast to proof objects -- $\lambda$ terms in an expressive type theory --  they combine full formality with a narrative (step-by-step with justifications) structure  conducive to human understanding.
Generally, proof presentation -- i.e. transforming proofs to mathematical vernacular -- has been studied in various contexts; see~\cite{AmerkadEtAl:ptp01,Hu-96-a,Horacek:tirdtm00} for details and pointers.

A particularly influential design has been the \textsf{ISAR}\footnote{inspired by the Mizar language, hence the name} (Intelligible semi-automated reasoning) proof document language~\cite{MW:IIAGFFHRPD} of the \textsf{Isabelle} proof assistant.
\textsf{ISAR} tries to bridge the semantic gap between prover-internal -- proof objects generated by tactic scripts --  and an appropriate level of abstraction for user-level work.
\textsf{ISAR} proof texts consist document constructors, atopic steps via high-level tactic invocations, and library references.
Thus they  admit a purely static reading, thus being intelligible later without requiring dynamic replay that is so typical for traditional proof scripts.
This is a very important characteristic of narrative representations. 

\newAspectFacet[N]{Semi-Formal Systems}S{Documents at the Semantic Level}

The semantic level relaxes the requirement of full formality and allows informal elements interspersed with formal ones.
For instance, \textsf{Isabelle} provides native syntax for {\LaTeX}-like commands as well as raw {\LaTeX}, and all \textsf{Isabelle} documents can be turned into {\LaTeX} for documentation.
Similarly, \textsf{Agda}~\cite{agda} can read two kinds of files: documentation files with interspersed \textsf{Agda} code or \textsf{Agda} code files with interspersed documentation.

Thus semiformal systems can choose which aspects to formalize and focus on services using those, leaving the informal ones to humans;
\cite{Kohlhase:tffm13} introduces the concept of \defemph{flexiformality} (flexible formality) and discusses the issues involved. 
A good example is \textsf{weak type theory}~\cite{KamNed:arbflm04}, a $\lambda$-calculus with a  linguistically inspired type system, which is intended as an intermediate step in the formalization of mathematical developments.
The \textsf{MathLang} system~\cite{KamWelZen:cmt14} based on ideas from \textsf{weak type theory} allows to annotate (i.e. flexiformalize) various aspects of a mathematical document, up to a point where enough semantic information to drive verification of the document in a proof assistant~\cite{Retel:CompVerifMathMathLangMizar09}.

The \textsf{OMDoc} (Open Mathematical Documents) format~\cite{Kohlhase:OMDoc1.2} is possibly the most complete framework for flexiformal mathematics.
It subsumes all the other representation formats and can serve as an interoperability layer. 
It specifies markup for mathematical documents and knowledge in a system-independent general framework and relates the two aspects.
\textsf{OMDoc} provides representational infrastructure at three levels: the object level for mathematical formulae, based on OpenMath\ednote{MK: introduce OpenMath somewhere, probably in the inference aspect} and content \mathml,  the statement level for definitions, theorems, and proofs, and the theory level for document sectioning and knowledge grouping.
\textsf{OMDoc} documents can be created by writing them in \sTeX~ \cite{Kohlhase:ulsmf08,sTeX:github:on}, a variant of {\LaTeX} that allows ``semantic preloading'', i.e. invisible \textsf{OMDoc} markup in the {\LaTeX} sources and then converting to \textsf{OMDoc/XML} via the {\LaTeX}ML system.
Conversely, \textsf{OMDoc} documents can be transformed into \defemph{active documents}, which use the semantic information for embedded services -- the more semantic preloading, the more services -- see~\cite{KohDavGin:psewads11}  for a discussion. 

Finally, another example of flexiformal -- here computational -- documents are \textsf{Wolfram Notebooks}~\cite{MathematicaNB:on} or \textsf{Jupyter}~\cite{jupyter-notebook:on} notebooks.
Here, computational ``cells'' with executable code (\textsf{Mathematica} for \textsf{Wolfram notebooks} and a wide variety of computational systems for \textsf{Jupyter notebooks}) are interleaved with text cells, which provide an (informal)  narrative.
Computational cells show the results of computations, the code can be arbitrarily edited and re-executed, giving a very flexible way of exploring the mathematical contents.
Computational cells can also drive ``widgets'', which pipe computation results into special-purpose interaction forms. 



\section{Primary Aspect: Organization}\label{pa:orga}
\ednote{responsible: Bill, Yasmine}
Every mathematical system organizes the body of mathematical knowledge (MK) relevant to
the system as a structure consisting of \emph{units of mathematical knowledge},
\emph{means for combining the units}, and an underlying \emph{semantics} for understanding
what the units and their combinations mean.

\newAspectFacet[O]{Compound Units of Mathematical Knowledge}C{}

There are several kinds of MK units, both \defemph{atomic} and \defemph{compound}.  Atomic
MK units are not composed of smaller MK units, while compound MK units contain components
that are MK units themselves. Examples of atomic units include equations, tables,
algorithms, definitions, theorems, and proofs, while examples of compound units include
theorem-proof pairs, term-definition pairs, question-answer pairs, articles, and axiomatic
theories.

\newAspectFacet[O]{Indexed Knowledge Collections}I{}

There are various ways that mathematical knowledge units can be combined in the structure.
The default case is that they are ordered in some kind of a \defemph{collection} and
indexed by some salient features (e.g. concept names or sizes) either as an organization
principle or for a retrieval-based user interface. A mere \emph{list} is a special case of
this, there items are indexed by their position, (mathematical) encyclopaedias like the
Encyclopedia of Mathematics~\cite{EncMath:on}, PlanetMath~\cite{planetmath:on},
Wikipedia~\cite{wikipedia:biblatex}, and Wolfram MathWorld~\cite{MathWorld:on} are prime
examples; they are indexed by concept names.  Other examples are mathematical data sets
like the \emph{Digital Library of Mathematical Functions} (DLMF)~\cite{DLMF} -- indexed by
function; the \emph{Online Encyclopedia of Integer Sequences} (OEIS)~\cite{OEIS:on}), and
triangle centers (as in the Encyclopedia of Triangle Centers (ETC)~\cite{ETC:on}.

An interesting case in this class is the MathOverflow~\cite{MathOverflow:on}, where
mathematical knowledge is organized (i.e.g indexed) by questions it answers. Of course
this poses interesting problems e.g. about when two questions are ``equal''; MathOverFlow
has developed community-organized solutions here.

\newAspectFacet[O]{Graph-Structured/Semantic Organization}G{}

In this more elaborate organizational principle, knowledge items are interconnected by
semantical relations, e.g. concepts in a taxonomy. Other examples of the latter include a
collection of articles connected by hyperlinks and a collection of axiomatic theories
interconnected by theory morphisms. The main mechanism of these graph-shape organization
forms (knowledge graphs) is that they support some kind of knowledge inheritance
mechanism, that enhances space-efficiency, consistency, and maintainability of the
knowledge collection.

Arguably the most general concept in this space is that of \defemph{theory graphs}: A
directed graph whose nodes are axiomatic theories and whose edges are theory morphisms.
The latter are meaning-preserving mappings that enable information to flow from abstract
theories to more concrete theories or equally abstract theories.  This organizational
structure is found in many proof assistants, specification systems, and logical
frameworks.

Collections of numeric or symbolic algorithms constitute an interesting instance of this
class of collections. They are usually organized by code/library dependencies, and thus
inherit functionality. Algorithm collections (we think of them as \defemph{algorithmic
  theories}) are embodied in computer algebra systems such as Maple~\cite{maple:on} and
Mathematica~\cite{mathematica:on}.

\newAspectFacet[O]{Heterogeneous Organization}H{} The units in an organization structure
can be \emph{homogeneous}, i.e., all of one kind, or \emph{heterogeneous}, i.e. of several
kinds.\ednote{MK@BF: we need to define this: do you mean Florian's ``homogenous
  vs. heterogeneous methods?}  For algorithmic theories, we think of a collection to be
heterogeneous, if it involves multiple programming languages or paradigms. For instance
the OEIS~\cite{OEIS:on} collects implementation of integer sequences in many computational
stystems.

Theorem Prover libraries, such as those of such as HOL Light~\cite{HOLLight-Library:on}
and Mizar~\cite{MizarKB:on} are heterogeneous: they are (essentially) lists of the axioms,
definitions, theorems; but they also compose a graph of deductive developments in which
the root of the tree is foundational axiomatic theory.

\newAspectFacet[O]{Formal Organization}F{}

The underlying semantics can be based on traditional informal
mathematical practice; formal logics like first-order logic, simple
type theory, set theory, and dependent type theory. 

\newAspectFacet[O]{Organization by Mathematical Practice}P{}

In contrast to the above, the knowledge units can be organized by mathematical
practice. An example is the Math Subject Classification~\cite{AMS:MSC2010}, an 

\begin{newpart}{MK: moved over from concretization, makes more sense here.}
  \paragraph{Explicitly represented mathematical knowledge}
  Most mathematical system represents mathematical knowledge. Either implicitly inscribed
  into the source code used in the implementation of the system or explicitly
  represented. Many systems manage quite a lot of represented knowledge and
  data.\ednote{MK: we need to talk about FAIR and what it means for math somewhere.}
  
  Figure~\ref{fig:datasets}\ednote{MK: we should probably extend this by some more systems} gives an overview over the scale of   mathematical knowledge
  explicitly represented in state of the art systems.
  We indicate the tetrapod aspects involved in the last column to give a preview o the
  discussions in Sections \ref{pa:inf} to \ref{nar}.
  Table \ref{tab:bigtable} below gives a more detailed classification of the aspects in
  terms of specific facets that will be introduced below.

\begin{figure}[htp]\centering\small
  \begin{tabular}{| p{.3\textwidth} | p{0.5\textwidth}|p{.1\textwidth}|}\hline
  Data/Knowledge in  & Description & Aspects\\\hline\hline
  Theorem prover libraries \cite{OAFproject:on}  & $\approx 5$ proof libraries, $\approx
                                                   10^5$ theorems each, $\approx 200$ GB &
    Inf\\\hline
  Computer algebra systems \cite{sagemath} & e.g., SageMath distribution bundles $\approx
                                             4$ GB of various tools and libraries& Comp\\\hline
  Modelica libraries \cite{Modelica:on} &$> 10$ official, $> 100$ open-source, $\approx 50$ commercial,
      $> 5.000$ classes in the Standard Library, industrial models can reach $.5$M
                                          equations & Comp\\\hline
 Integer Sequences \cite{OEIS:on} & $\approx 330$K sequences, $\approx 1$ TB  & Conc, Inf,
    Nar\\\hline
 Sequence Identities \cite{kwarc:datahost:on} & $\approx .3$M sequence identities,
                                                $\approx 2.5$ TB & Comp, Inf\\\hline
 Highly symmetric graphs, maps, polytopes \cite{ConderCensuses:on, HartleyPolytopes:on,
    LeemansPolytopes:on, PotocnikCensuses:on, RoyleVT:on, WilsonET:on} &
                                                                         $\approx 30$
                                                                         datasets,
                                                                         $\approx
                                                                         2\cdot10^6$
                                                                         objects, $\approx
                                                                         1$ TB & Conc\\\hline
  Finite lattices \cite{KohLat:on, LeeLat:on, MalLat:on} & $7$ datasets, $\approx 17 \cdot
                                                           10^9$ objects, $\approx 1.5$ TB
                                   & Conc\\\hline
  Combinatorial statistics and maps \cite{findstat} & $\approx1.500$ objects& Conc \\\hline
  SageMath databases \cite{SageDB:on} & $12$ datasets & Comp, Comp\\\hline
  $L$-functions and modular forms \cite{lmfdb:on} & $\approx 80$ datasets, $\approx 10^9$
                                                    objects, $\approx 1$ TB & Comp, Conc, Nar\\\hline
   zbMATH \cite{zbMATH:on} & $\approx 4$M publication records with semantic data, $\approx
                             30$M reference data, $>1$M disambig. authors, $\approx 2,7$M
                             full text links: $\approx 1$M OA & Nar, Link, \\\hline
  swMATH \cite{swMATH:on} & $\approx 25$K software records with $> 300$K links to $> 180$K
                            publications& Nar,Link  \\\hline
  EuDML  \cite{EuDML:on} & $\approx 260$K open full-text publications& Nar \\\hline
  Wikidata  \cite{wikidata:on} & $34$ GB linked data, thereof about $4$K formula entities,
                                 interlinked, e.g., with named theorems, persons, and/or
                                 publications & Link, Nar \\\hline
  arXiv.org & $\approx 300$K math preprints (of $\approx1.6$M) most with {\LaTeX} sources&
    Nar,link\\\hline
  MathOverFlow & $\approx 1,1$M questions/answers, $\geq11$K answer authors & Nar\\\hline
  Stacks project & $\geq 6000$ pages, semantically annotated, curated, searchable textbook
                                   & Nar\\\hline
  nLab & $\geq 13$K pages on category theory and applications& Nar\\\hline
\end{tabular}
  \caption{Represented Knowledge/Data in Mathematical Software Systems}\label{fig:datasets}
\end{figure}
\end{newpart}



\section{The Big Table of Systems and their Aspects}
\begin{table}[ht]\centering\footnotesize
  \begin{tabular}{|l|p{2.4cm}|p{2.5cm}|l|l|l|l|l|}\hline
    & System & Reference & Orga. & Inference & Comp. & Concr. & Narration\\\hline\hline
    \multirow{8}{*}{\begin{sideways}Deduction\end{sideways}}
    & Coq & \cite{coq} & & C A T P R & & &  \\\cline{2-8} 
    & Isabelle &\cite{isabelle:on} & & P & & &S \\\cline{2-8} 
    & Mizar & \cite{mizar:online} & & P & & &\\\cline{2-8} 
    & Otter & \cite{McCune:otter03} & & A & & &\\\cline{2-8} 
    & * Hammer & & & A & & &\\\cline{2-8} 
    & CVC/Z3 & & & & & &\\\cline{2-8} 
    & Prolog & \cite{ClMe81} & & L & & &\\\cline{2-8} 
    & FOIL & \cite{Quinlan:lldfr90} & & L & & &\\\hline 
    \hline
    \multirow{9}{*}{\begin{sideways}Computation\end{sideways}} 
    & Mathematica & \cite{mathematica:on} & & & & &S\\\cline{2-8} 
    & SageMath & \cite{SageMath:on} & & & & & \\\cline{2-8} 
    & GAP & \cite{GAP:on} & & & & & \\\cline{2-8} 
    & GeoGebra & \cite{geogebra:on} & & & & & \\\cline{2-8} 
    & CoCalc &\cite{cocalc:on} & & & & & \\\cline{2-8} 
    & Octave & \cite{octave:on} & & & & & \\\cline{2-8} 
    & Simulink & \cite{simulink:on} & & & & & \\\cline{2-8} 
    & R & \cite{R:on} & & & & & \\\cline{2-8} 
    & Stan & \cite{stan:on} & & & & & \\\hline 
    \hline
    \multirow{7}{*}{\begin{sideways}Concretization\end{sideways}} 
    & Math Gene. Proj.& \cite{MGP:on} & & & & \AF{CL} & \\\cline{2-8} 
    & WikiData &\cite{wikidata:on} & K & & & \AF{CR} \AF{CL} & \\\cline{2-8} 
    & OEIS & \cite{OEIS:on} & E & & & \AF{CR} & \AF{NP} \\\cline{2-8} 
    & LMFDB & \cite{Cremona:LMFDB16,lmfdb:on}& & & & \AF{CR} & \\\cline{2-8} 
    & Small Groups Lib. & \cite{GapSmallGroups:on} & & & & \AF{CR} \AF{CC}\cm{GAP} \AF{CE}& \\\cline{2-8} 
    & DLMF & \cite{DLMF,Loz:DLMF} & & & & \AF{TS} & \AF{NS}\\\cline{2-8} 
    & Inv. Symb. Calc. & & & & & \AF{TS} & \\\hline 
    \hline
    \multirow{8}{*}{\begin{sideways}Narration\end{sideways}}
    & arXiv & \cite{arxiv:online} & & & & & \AF{NP} \\\cline{2-8} 
    & zbMath & \cite{zbMATH:on} & & & & \AF{CL} & \AF{NP}\\\cline{2-8} 
    & \TeX/\LaTeX & \cite{Knuth:ttb84,Lamport:ladps94}& & &?\cm{tex} & & \AF{NP} \AF{NS}\cmref{tex}\\\cline{2-8} 
    & p\mathml & \cite{CarlisleEd:MathML3:on}& & & & & \AF{NP} \\\cline{2-8} 
    & \sTeX & \cite{Kohlhase:ulsmf08} & & & & & \AF{NP} \AF{NS} \AF{NF}\\\cline{2-8} 
    & SIUnitsX & \cite{siunitx:on}& & & & & \AF{NF}\cmref{units} \\\cline{2-8} 
    & OpenMath/ c\mathml & \cite{BusCapCar:2oms04,CarlisleEd:MathML3:on}& & & & & \AF{NF}\cm{om} \\\cline{2-8} 
    & Wikipedia & \cite{wikipedia:biblatex} & E\cm{wikia} & & & & \AF{NP} \AF{NS}\cm{wikib}\\\hline\hline 
    \multirow{6}{*}{\begin{sideways}Orga.\end{sideways}} 
    & MathOverflow & \cite{MathOverflow:on}& \AF{OI} & & & & \AF{NP}\\\cline{2-8} 
    & Polymath & \cite{PolymathBlog} & & & & & \AF{NP} \\\cline{2-8} 
    & AFP & \cite{AFP:online} & \AF{OG} \AF{OH} & & & & \\\cline{2-8} 
    & MSC & \cite{MSC2010} & \AF{OP} & & & & \\\cline{2-8} 
    & MathHub & \cite{MathHub:on} &\AF{OG} \AF{OG}&&&&\\\hline
    \hline
  \end{tabular}
  \caption{Tetrapod Systems and their Aspects}\label{tab:bigtable}
\end{table}
\ednote{MK: I had problems finding or citing the inverse symbolic calculator. There does not seem to be a working version online. Do we care?}
\ct{tex}{\TeX~\cite{Knuth:ttb84} pairs a set of layout primitives with a Turing-complete macro expansion facility, which is use d by a large community to define libraries of macros.  \LaTeX~\cite{Lamport:ladps94} is the widely used one; it establishes semantic markup for sectioning, crossreferences, bibliographic references, and statements.}
\ct{units}{Additional packages can be used to extend semantic markup, e.g. the \textsf{SIUnitsx} package for quantity expressions.}
\ct{om}{\textsf{OpenMath} and -- by reference -- \textsf{presentation} \mathml (whose semantics is given in terms fo \textsf{OpenMath}) fully describe the structure of mathematical formulae and give the meaning of the symbols in terms of \textsf{OpenMath Content Dictionaries}, which are semantic mathematical documents themselves.}
\ct{wikia}{Wikipedia uses a restricted subset of \TeX to create \textsf{presentation} \mathml.}
\ct{wikib}{Some of the links and concepts are classified with Linked Open Data annotations.}
\ct{GAP}{The Small Groups library is deeply integrated into the GAP computer algebra system and uses the GAP-internal representations to represent the groups extremely space-efficiently (a handful of bits per group). In particular, extracting the data out of GAP leads to a dataset that is multiple orders of  magnitude less space-efficient.} 
\ednote{Conjecturing, Theory Exploration}

\ednote{not Python}
\ednote{Matlab, Wolfram Alpha? Geogebra?}
\ednote{Geogebra represents Graphing Calc}
\ednote{Numeric/Scientific computation, Optimisation, Statistics: refer to surveys?}
\ednote{Keep in mind: IDRIS, F*}
\ednote{Otter representative of Prover 9, Mace}
\ednote{probabilistic proof? is it a computation thing?}
\ednote{zbMath representative for MathSciNet, has swMath, Google Scholar, Scopus, arXiv}
\ednote{MathTutor History of Mathematics (bibliographies) archive http://www-history.mcs.st-and.ac.uk/}
\ednote{not Oracle}
\ednote{not InDesign, Markdown}
\ednote{MathWorld}

We give a overview of mathematical software systems from a tetrapod perspective in Table ~\ref{tab:bigtable}. We list systems in the first row, and specify which aspects they support in the last five using the letter codes specified in the ``sub-aspects'' in Sections~\ref{pa:orga} to \ref{pa:nar}. Note that the letter codes are only unique per column.\ednote{Make an example where this happens} Where necessary, we mark the codes with comments which can be referenced in the list below. 

\cmlist


\section{Realizing Secondary Aspects}
\newcommand{\lb}{\linebreak}

\begin{figure}\footnotesize\centering
\begin{tabular}{|l||p{2.2cm}|p{2.2cm}|p{2.2cm}|p{2.2cm}|p{2.2cm}|}
\hline
Primary & \multicolumn{5}{c|}{Secondary Aspect} \\
Aspect & Org & Inf & Comp & Tab & Narr \\
\hline
\hline
Org  & category of theories \lb module systems
     & type inference
     &
     &
     & \\
\hline
Inf  & theories
     & meta-theorems \lb verification \lb proof-checking
     & recursion \lb rewriting \lb logic programming \lb tactics \lb ATP \lb decision procedures
     &
     & documentation \lb semi-formal proofs\\
\hline
Comp & specifications
     & verification
     & preprocessing \lb code generation \lb profiling
     & memoization \lb package repositories
     & documentation \\
\hline
Tab  & schemas
     & 
     & querying \lb built-in functions
     & 
     &\\
\hline
Narr &
     &
     & active documents \lb literate programming \lb macros
     &
     & documentation \\\hline
\end{tabular}
\caption{Paradigms for Supporting Secondary Aspects}\label{fig:support}
\end{figure}

Systems usually use additional aspects, which we call secondary.
The secondary aspect can have multiple roles:
\begin{compactitem}
 \item It might enhance knowledge written in the primary aspects, e.g., narrative documentation of programs.
 \item It may substitute for knowledge written in the primary aspects, e.g., a narrative snippet describing an omitted proof step in a semi-formal proof.
 \item It may be used to talk about knowledge written in the primary aspect, e.g., a computational tactic that produces proofs or the verification of a computational system.
\end{compactitem}

Note that some of these roles allow for the primary and secondary aspect to be the same.
For example, we can use a computational preprocessor to generate programs before compilation.
Or we can proof a meta-theorem that states the admissibility of an additional inference rule.


\section{Conclusion}
In this living survey we survey the state of the art in mathematical
software systems from the perspective of the tetrapodal model of mathematical
knowledge introduced in~\cite{CarFarKohRab:bmobb19}.  Other than the survey
itself, we also contribute a set of facets of the main five aspects
\begin{inparaenum}[\em i\/\rm)]
\item inference,
\item computation,
\item concretization,
\item narration, and
\item organization
\end{inparaenum}
of the tetrapod (see Figure~\ref{fig:tetrapod}). Each paradigmatic
mathematical software system is analyzed with these facets in mind---
with the results in Table~\ref{tab:bigtable}.  The development of the facets
has been an iterative process of recognizing patterns
in system functionality, using these for classification, and verifying
the soundness of the results on the systems.  The table shows that current
systems are still predominantly single-aspect, but some trans-aspect
facets are creeping in. We hope that this trend continues, and that we will see
true tetrapodal systems in the future.

We will monitor the situation in later versions of the survey.


\printbibliography

\end{document}

\bibliography{../macros/fr/bib/pub_rabe,../macros/fr/bib/rabe.bib,../macros/fr/bib/systems.bib}

